\documentclass[11pt,twoside]{article}

\usepackage{asp2006}
\usepackage{epsf}
\usepackage{psfig}
\usepackage{lscape}
\usepackage{graphicx}

\begin{document}
\title{Hypervelocity Stars: Young and Heavy or Old and Light?}
\author{Uli Heber,$^1$ Heiko Hirsch,$^1$ Heinz Edelmann,$^{1,6}$ Ralf 
Napiwotzki,$^2$ Simon O`Toole,$^3$ Warren Brown,$^4$ and Martin Altmann$^5$}
\affil{$^1$Dr. Remeis-Sternwarte, Universit\"at Erlangen-N\"urnberg, 
Sternwartstr.~7, D-96049 Bamberg, Germany}
\affil{$^2$Centre for Astrophysics Research, University of Hertfordshire,
             College Lane, Hatfield AL10 9AB, UK}
\affil{$^3$Anglo-Australian Observatory, PO Box 296 Epping, NSW 1710, Australia}
\affil{$^4$Smithsonian Astrophysical Observatory, 60 Garden St, Cambridge, 
MA~02138, USA}
\affil{$^5$Astronomisches Rechen-Institut,    
M\"onchhofstr.~12-14, D-69120 Heidelberg,
Germany}
\affil{$^6$McDonald Observatory, University of Texas at Austin, 1 University Station, C1402, Austin, 
TX 78712-0259, USA}

\begin{abstract} 
The first three hyper-velocity stars (HVS)
unbound to the Galaxy were serendipitously discovered in
2005. The only suggested origin of hyper-velocity stars 
is the Galactic Centre as it hosts a super-massive black hole capable of
accelerating stars to such high velocities.  
Only one HVS, the sdO star US~708, is known to be an old low mass star, 
while HE~0437$-$5439 is an apparently normal early-type B-star, too short-lived
to originate from the Galactic Centre, but could possibly come from the LMC. 
A systematic survey has led to the
discovery of seven new HVS of late B-type (similar to the prototype HVS1), 
which can
either be massive stars ($\approx 3~M_\odot$) or horizontal branch stars,
sufficiently long-lived to have travelled from the Galactic Centre. 
We present new spectral analyses of five known HVS as well as of a newly
discovered candidate. It is possible that the late B-type HVS are a mix of 
main sequence and
evolved BHB stars. In view of the time scale problem we revisit 
HE~0437$-$5439 and discuss a possible subluminous nature of this star. 
\end{abstract}


\section{Introduction}   
Stars moving at velocities higher than the Galactic escape velocity were first
predicted to exist by \cite{t15_hills88}. The first such hyper-velocity stars (HVS) 
were discovered serendipitously only recently 
\citep{t15_brown05, t15_hirsch05, t15_edelmann05}.     
A systematic search for such objects has resulted in the discovery of seven
additional HVS up to now \citep{t15_brown06a, t15_brown06b, t15_brown07}.

\cite{t15_hills88} predicted that the tidal disruption of a binary by an
super-massive black hole (SMBH) could 
lead to the ejection of 
stars with velocities exceeding the escape velocity of our Galaxy. The Galactic
Centre (GC) is the suspected place of origin of the HVSs as it hosts an SMBH.
\cite{t15_yu03} investigated Hill's SMBH slingshot mechanism further and estimate a 
HVS formation rate of $10^{-5}\,{\rm yr^{-1}}$ which 
implies $\approx$\,2\,000 HVS in a sphere 
of $120\,{\rm kpc}$ radius \citep{t15_brown06a}. 
If the GC hosts a tight binary of a 
SMBH and an intermediate black hole (IMBH), the formation rate for 
HVS would be ten times as large \citep{t15_yu03}. 

\cite{t15_gnedin05} and \cite{t15_bromley06} showed that the space 
distribution of HVSs provides significant constraints on the 
shape and density distribution of the Galactic dark matter halo.
\cite{t15_brown07} estimate that 100 HVS are possibly enough to constrain
ejection mechanisms and (dark matter) potential models for the Galaxy. 

In view of the great importance of hyper-velocity stars to astrophysics it is
worthwhile to study each of the known HVS in as much detail as possible.
As the stars are found at high Galactic latitudes one might expect them to be
old, low mass stars. However, there is evidence to the opposite as HVS1 and 
HE~0437$-$5439 show photometric and spectroscopic signatures of young massive
stars. 
US~708 (HVS2) is the only bona-fide old, low mass star amongst the
known HVS because it is a very hot helium-rich sdO star. 
All other HVS discovered are 
either late B-type dwarf 
stars (of about 3 $M{_\odot}$) or blue horizontal branch stars. Additional
observational evidence is lacking up to now to distinguish between these
options.  

In the context of this conference we shall discuss whether some of the HVS stars
(besides US~708) could be related to the (extreme) horizontal branch.    
We begin with a description of the hyper-velocity sdO star and present
preliminary results of spectral analyses of HVS of late B type in Section~3. 
Section~4 presents speculations about 
the early B-type star HE~0437$-$5439 for which a LMC origin was suggested. 
Before concluding we present
a preliminary spectral analysis of a candidate HVS, a bright B-type
giant, in 
Section~5.


\section{US~708 -- A Hyper-Velocity Subdwarf O Star}  

\begin{figure}
\centerline{\includegraphics[width=0.6\textwidth]{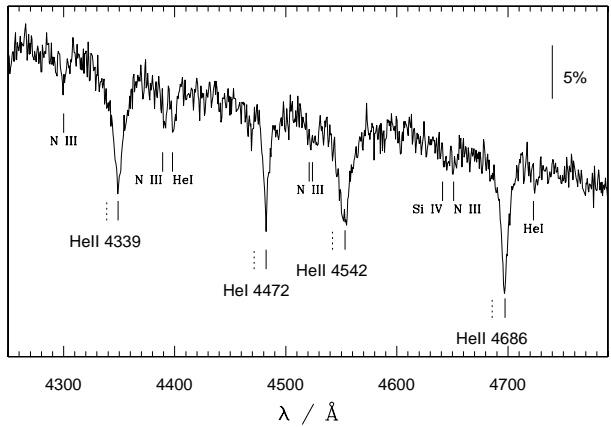}}
\caption{Section of the spectrum of US~708. Rest-wavelengths of the strongest 
lines are marked as dashed lines. Note the large redshifts \citep[from
][]{t15_hirsch05}.}\label{t15_us708_keck}
\end{figure}

Amongst sdO stars drawn from the Sloan Digital Sky Survey,   
\cite{t15_hirsch05} discovered a hyper-velocity star,
US~708, with a heliocentric radial velocity of 
+708$\pm$15~$\mathrm{km\,s^{-1}}$. 
A quantitative NLTE model atmosphere analysis of optical spectra obtained 
with the KECK I telescope (see Fig.~\ref{t15_us708_keck})
shows that US~708 is a \emph{helium-enriched} sdO 
and that its atmospheric parameters ($T_{\rm eff}$=44\,500~K, $\log(g)=5.25$)
are typical for this spectral class \citep{t15_stroer07}.
 Adopting the canonical mass of
half a solar mass from evolution theory the corresponding distance is 19~kpc. 
Its galactic
rest frame velocity is at least 757~$\mathrm{km\,s^{-1}}$, 
much higher than the local Galactic escape velocity (about
430~$\mathrm{km\,s^{-1}}$)
indicating that the star is unbound to the Galaxy.

Numerical kinematical experiments
were carried out to reconstruct the path of US~708 from the GC. 
US~708 needs about 36~Myr to travel from the GC to its present
position, which is shorter than its evolutionary lifetime. Hence it is plausible
that the star might have originated from the GC, which can be tested by
measuring accurate proper motions.

\section{Hyper-Velocity Stars: Massive or Low Mass?}

\begin{figure}
\centerline{\includegraphics[width=0.65\textwidth]{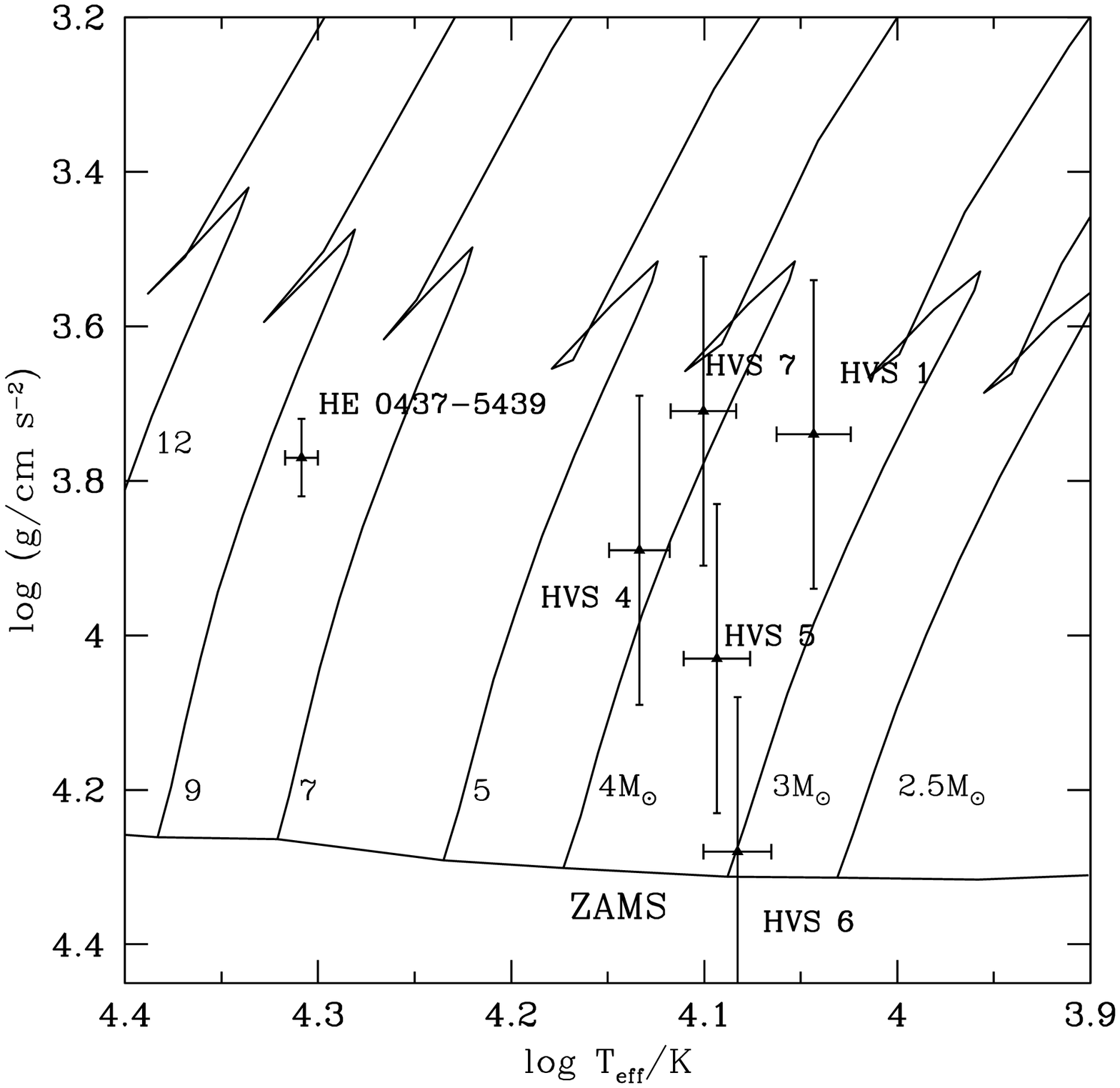}}
\centerline{\includegraphics[width=0.65\textwidth]{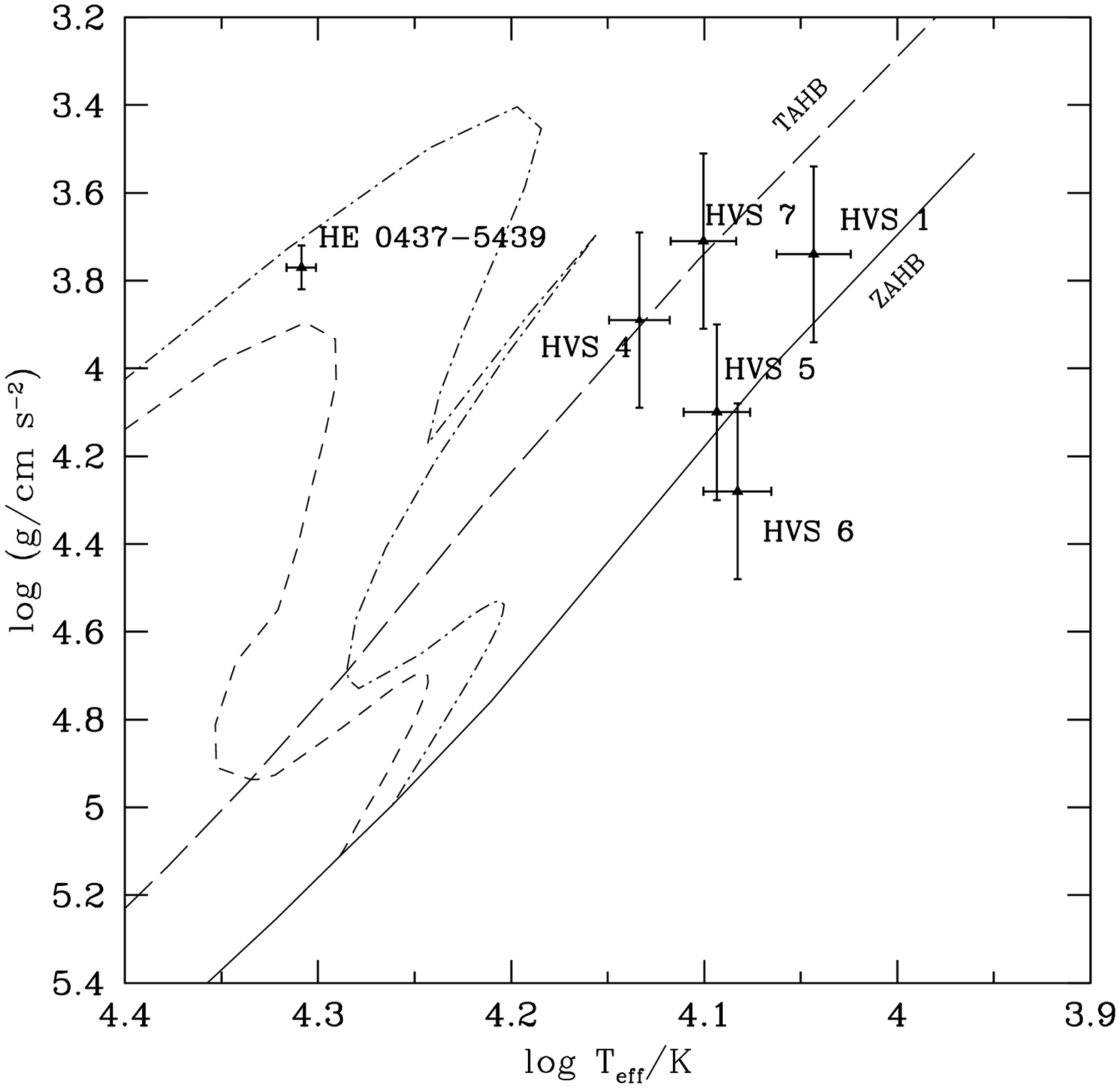}}
\caption{Position of the B-type HVSs (including HE~0437$-$5439) in a 
($T_{\rm eff}$, $\log(g)$) diagram
and comparison with evolutionary models and tracks to determine 
masses and evolutionary ages. The atmospheric parameters were derived
assuming zero rotation velocity and solar helium abundance 
except for HVS~1.
{\it Top panel}: comparison with models for massive stars from \citet{t15_schaller92}
{\it Lower panel}: comparison with models for the 
Horizontal Branch. The
ZAHB is the zero age horizontal branch, while TAHB (dashed) 
is the terminal age horizontal branch. 
Post-EHB tracks for $0.485 M_\odot$ (short-dashed) and $0.49 M_\odot$ (dashed
dotted) are from \cite{t15_dorman93}.}\label{t15_hvs_mass}
\end{figure}

The hyper-velocity stars discovered in the systematic survey of 
\cite{t15_brown06a,t15_brown06b,t15_brown07} 
are all
of late B spectral type. As the location of the main sequence intersects with
that of the horizontal branch in the HRD near this spectral type, it is not 
clear a priori whether
the stars are un-evolved massive stars of about 3 $M_\odot$ or evolved low mass 
stars of about half a solar mass. Hence we lack crucial information on their 
distances which is important for an 
analysis of their kinematics. 

Medium resolution (1.2\AA) 
MMT spectra of the HVS stars 1,4,5,6, and 7 were obtained by Brown and the
observed Balmer lines were matched 
to a grid of
synthetic spectra calculated from metal-line blanketed LTE model atmospheres 
\citep[see][]{t15_heber00,t15_otoole06} assuming solar helium abundance.
The helium line 4026\AA\ was included in the fit, but turned out to be too weak
to reliably constrain the helium abundance in most cases. 
Preliminary results are shown in 
Fig.~\ref{t15_hvs_mass}.     

From their position in the ($T_{\rm eff}$, $\log g$-) diagram it is evident 
that they can either be distant (intrinsically bright) main sequence B
(Fig.~\ref{t15_hvs_mass}, upper panel) or closer (intrinsically faint) blue horizontal branch 
stars (BHB, Fig.~\ref{t15_hvs_mass}, lower panel).
Obviously, the available information ($T_{\rm eff}$, $\log g$) is insufficient 
to distinguish a BHB star from a main sequence B star.

Hence, we inspected the line profiles in more detail. For HVS~1 the observed
profiles are broader than predicted by the synthetic spectra, which we attribute
to rotation. By varying the projected rotational velocity we find 
$v \sin i=190\mathrm{km\,s^{-1}}$ to fit best. There is also
evidence for rotational broadening in HVS~6. Such high projected rotation
velocities would favour a main sequence nature of the stars. 
Photometric investigations revealed HVS~1
to be a slowly pulsating B-type main sequence stars  
\citep{t15_fuentes05}. 
The helium abundance of HVS~7 is subsolar 
(almost by a factor of one hundred) which is typical for a horizontal branch star but
rather unusual for a main sequence star.
Thus it is possible that the late B-type HVS are a mix of main sequence and
evolved BHB stars.



\section{HE~0437$-$5439  Revisited}

The late B type stars as well as the 
sdO star are sufficiently long-lived ($\approx$ 100~Myr) to have reached
their present location in the Galactic halo after ejection from the GC.  
The third HVS, HE~0437$-$5439 \citep{t15_edelmann05} is much shorter-lived as it is of  
early B-type. $T_{\rm eff}$=20\,350~K,
$\log(g)=3.77$, and a solar helium abundance were derived by a quantitative
analysis of high-resolution spectra.
Solar abundances are consistent with the observations to within a factor of 
two to three.  
Both, the chemical composition and the
moderate rotational velocity ($v\sin(i)=54~\mathrm{km\,s^{-1}}$) of HE~0437$-$5439 were
considered evidence
for a main sequence nature. Accordingly, the star lies at similar
distance (60~kpc) as the other HVSs, but is of much higher mass ($8~M_{\sun}$).

Numerical kinematical experiments were carried out to trace the trajectory 
of HE~0437$-$5439 from the Galactic Centre to its present location in the Galactic halo. 
However, its travel time (100~Myr)
was found to be much longer than its main sequence lifetime ($\approx$ 
25~Myr) rendering a GC origin unlikely.  
\cite{t15_edelmann05} suggested that 
the star could have originated from the Large Magellanic Cloud 
(LMC) as it is much closer to it (18~kpc) than to the GC. Indeed HE~0437$-$5439  can
reach its position from the centre of the LMC in less than its life time if its
ejection velocity were of the order of 500$\mathrm{km\,s^{-1}}$ or larger. 
However, no SMBH is known to exist in the LMC.

As the focus of this conference lies with hot subluminous stars we feel free to
discuss a somewhat speculative solution 
to the time scale problem. Let us assume that HE~0437$-$5439 is not 
a young massive star 
at all, but merely mimics it very closely. 
Comparing the position of HE~0437$-$5439  in the 
($T_{\rm eff}$, $\log g$) diagram
 to evolutionary tracks (Fig. \ref{t15_hvs_mass}) we
conclude that  HE~0437$-$5439 could be a post-HB star of $\approx 0.5 M_\odot$. 
Usually, blue horizontal branch stars in the halo of our Galaxy show peculiar 
abundance pattern and very low projected rotation velocity 
($\le 10\mathrm{km\,s^{-1}}$ ) quite different from those of early type 
MS stars.
So, how could a BHB star maintain (i) a high rotational velocity as well as (ii)
normal abundances? (i) If the SMBH slingshot mechanism is valid, the star was in a
binary before ejection. In that system the rotation would have been tidally 
locked to
the orbit, enforcing a high rotation rate if the binary system was sufficiently
close. (ii) The abundance peculiarities in BHB stars are due to atmospheric
diffusion processes. To maintain normal abundances diffusion needs to be
suppressed. We conjecture that rotational induced motions in the atmospheres
could wash out the slow diffusive motions.
The rapid rotation enforced by tidally locking could then inhibit abundances 
peculiarities to built up. 
Hence tidally locked rotation and suppression of diffusion could result
in a spectrum of a BHB star that mimics that of a
massive star quite closely.  

If we assume a mass of 0.5 $M_\odot$ the star would be much 
much closer ($d$=14kpc).
Ejected at 927$\mathrm{km\,s^{-1}}$ the 
proper motion
required would be sufficiently large ($\mu_\alpha \cos{\delta} = 3.085$\,mas/yr and 
$\mu_\delta = -3.365$\,mas/yr) to be measurable with present day 
instrumentation.
The travel time from the GC (26~Myr) is much shorter than for a main sequence
star. However the evolutionary life time for the post-HB phase is also much
shorter ($\approx 10^7$ yrs). Hence the star has to have 
evolved from a more compact ($R=0.2 R_\odot$) extreme horizontal 
branch star, i.e.\ an sdB star (see Fig.~\ref{t15_hvs_mass}, lower panel) after ejection from
the GC. This poses an angular momentum problem, if angular momentum is conserved 
during evolution from the HB. The rotation velocity of the sdB progenitor 
of HE~0437$-$5439 would then be larger than
300$\mathrm{km\,s^{-1}}$. 
This implies that the orbital period of the binary before disruption had to be 
45 min.\ or less. If HE~0437$-$5439  was ejected by the disruption of an sdB 
binary due to tidal interaction with the SMBH in the GC, the original companion would 
have to be very compact, possibly a neutron stars or black hole. 

All in all this scenario for the origin of HE~0437$-$5439 appears to be 
far-fetched and the origin of the star as a massive one ejected from the LMC is
more plausible. An accurate proper motion
measurement would allow us to distinguish between the LMC
origin as a massive star or the GC origin as a low mass star. 


\section{HD~271791 -- a B-type Giant at High Velocity}

\begin{figure}
\centerline{\includegraphics[width=0.55\textwidth]{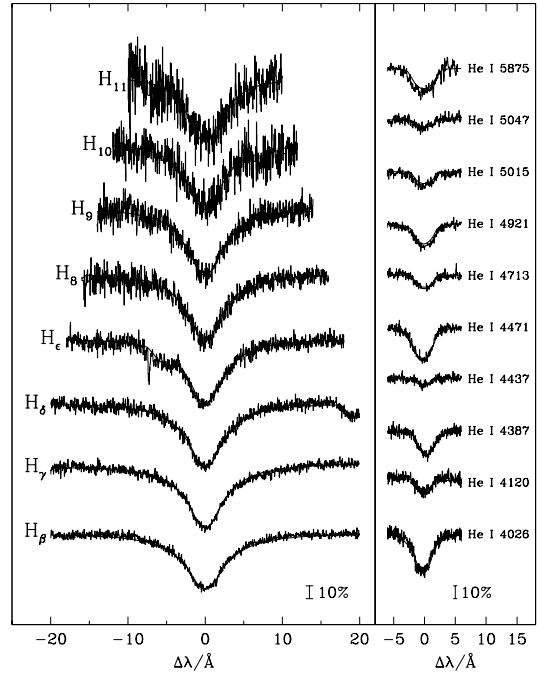}}
\caption{LTE line profile fit of the FEROS spectrum of HD~271791}\label{t15_fit_om88}
\end{figure}

\begin{figure}
\centerline{\includegraphics[width=0.55\textwidth,angle=270]{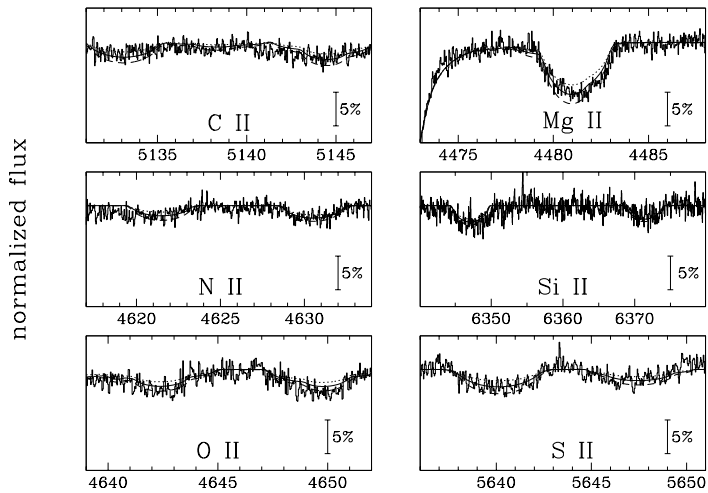}}
\caption{Selected metal line profiles in the FEROS spectrum of HD~271791 
compared to synthetic spectra with solar metal abundance (solid lines),
0.5 times solar metal abundance (dotted lines), 
and twice solar metal abundance (dashed lines).}\label{t15_metals_om88}
\end{figure}

In the course of our investigations of B-type stars at high Galactic latitudes 
we took high resolution
spectra of the bright star HD~271791 ($V=12.3$) with the ESO 2.2m telescope and 
the FEROS spectrograph yielding a heliocentric radial velocity of 
441$\mathrm{km\,s^{-1}}$.
We determine the atmospheric parameters of HD~271791 to be
$T_{\rm eff}=17810\pm 180$~K, $\log(g)=3.04\pm 0.03$~(cgs), and
$\log(n_{\rm He}/n_{\rm H})=-0.81\pm 0.02$.
The atmospheric parameters places HD~271791 within
the domain of B-type giants.
The projected rotational velocity is 
determined by a $\chi^2$ fit to all Balmer and helium lines.
The best matching fit results in a $v\sin(i)=124~\mathrm{km\,s^{-1}}$ (see Fig.
\ref{t15_fit_om88}).
Metal absorption lines of
\ion{C}{ii}, 
\ion{N}{ii}, 
\ion{O}{i},
\ion{O}{ii}, 
\ion{Ne}{i},     
\ion{Mg}{ii}, 
\ion{Al}{iii},  
\ion{Si}{ii},
\ion{Si}{iii}, 
\ion{S}{ii},
and
\ion{S}{iii}  
are clearly present in the FEROS spectrum.
However, all lines are broadened due to the high rotation.
This renders a quantitative abundance analysis difficult. 
Nevertheless, we compared synthetic spectra calculated 
from LTE model atmospheres \citep{t15_heber00} 
with different metal contents
to the FEROS spectrum (see Fig. \ref{t15_metals_om88}) and find that solar composition
provides a good match to the observations. 


The almost solar metal content and its high rotation suggests that 
HD~271791 is a young massive star.
Comparing its position in the 
($T_{\rm eff}$, $\log g$)
diagram to evolutionary tracks \citep{t15_schaller92}, 
a mass of $11.5\pm0.5 M_{\odot}$ and a very short 
evolutionary time $T_{\rm evol}\approx 17$~Myr result. 
Using the mass, effective temperature, gravity, and apparent
magnitude 
we derive a distance of $d=24$~kpc.

Unlike for the other HVS stars, proper motion measurements are available for
 HD~271791 and have been published in five astrometric catalogues.  
These values diverge
quite significantly, however, most showing a very small proper motion in 
right ascension and a substantial one in declination. Using the measurements
that roughly agree with each other, we derived a ''best'' value of 
$\mu_\alpha=-1$~mas/yr and $\mu_\delta=7$~mas/yr. Because of the large distance 
the corresponding transversal velocities are high resulting in 
galactic rest-frame velocities between 413~$\mathrm{km\,s^{-1}}$ and 
1080~$\mathrm{km\,s^{-1}}$ for different choices of the proper motion.

Numerical kinematical experiments were performed to trace the motion of the star
during its life time.
None of the trajectories took the star anywhere near the 
Galactic Centre. The peri-galactic distances vary mostly
between 15 and 20~kpc.
Therefore, it unlikely that this objects was born near the Galactic
Centre. 
Hence, we encounter the same time scale problem as for HE~0437$-$5439. Again 
we
might start to speculate about a low mass nature of HD~271791. 
Due to its 
low gravity the star would probably be in the post-AGB phase of evolution.
The post-AGB phase is very short-lived and thus unlikely to be 
observed.
Hence we regard the low mass option for HD~271791 very unlikely.


\section{Conclusions}

Only one HVS, the sdO star US~708 is an old low mass stars, while two stars 
are apparently normal early-type B-stars, too short-lived to have travelled from the 
GC to their present positions in the Galactic halo.
Seven newly discovered HVS are of late B-type (similar to the prototype HVS1), 
which can
either be massive stars ($\approx 3\,M_\odot$) or horizontal branch stars,
sufficiently long-lived to have travelled from the Galactic Centre. 
We presented new spectral analyses of five known HVS as well as of the newly
discovered candidate HD~271791, an apparently normal B giant. We find 
evidence for rapid rotation in HVS1 \& 6, and HD~271791, suggesting that 
they are massive stars. It may be possible that the late B-type HVS are a mix of main sequence and
evolved BHB stars. 

In view of the time scale problem we revisited 
HE~0437$-$5439 and discussed whether it could possibly be a subluminous star, which we,
however, 
find unlikely because of angular momentum constraints. Accurate proper motion
measurements of HE~0437$-$5439 are required to distinguish between a LMC
origin as a massive star or a Galactic origin as a low mass star. 
A NLTE analysis of high-resolution spectra is underway to determine
accurate abundances (Przybilla et al.,\,in prep.).
Alternative
ejection scenarios need to be looked at that do not require an SMBH 
\citep{t15_gvar07} 
or a much less massive IMBH \citep[$>1000\,M_\odot,$][]{t15_gual07}.



\begin{thebibliography}{}

\bibitem[Bromley et al.(2006)]{t15_bromley06} Bromley, B. C., Kenyon, S. J., 
Geller, M. J. et al. 2006, 
ApJ 653, 1194 
\bibitem[Brown et al.(2005)]{t15_brown05} Brown, W. R., Geller, M. J., Kenyon, 
S.J., \& Kurtz, M. J. 2005, ApJ 622,
L33

\bibitem[Brown et al.(2006a)]{t15_brown06a} Brown, W. R., Geller, M. J., Kenyon, 
S. J., \& Kurtz, M. J. 2006a, ApJ 640, 
L35

\bibitem[Brown et al.(2006b)]{t15_brown06b} Brown, W. R., Geller, M. J., 
Kenyon, S. J., \& Kurtz, M. J.
2006b, ApJ 647, 303 

\bibitem[Brown et al.(2007)]{t15_brown07} Brown, W.~R., Geller, 
M.~J., Kenyon, S.~J., et al.\ 2007, arXiv:0709.1471 


\bibitem[Dorman et al.(1993)]{t15_dorman93} Dorman, B., Rood, R.~T., 
\& O'Connell, R.~W.\ 1993, \apj, 419, 596 


\bibitem[Fuentes et al.(2005)]{t15_fuentes05}Fuentes, C.~I., Stanek, K.~Z., 
Gaudi, B.~S., et al.\ 2005, 

\bibitem[Edelmann et al.(2005)]{t15_edelmann05}
 Edelmann, H., Napiwotzki, R., Heber, U., et al.\ 
 2005, \apj, 634, L181


\bibitem[Gnedin et al.(2005)]{t15_gnedin05} Gnedin, O. Y., Gould, A., 
Miralda-Escude, J., \& Zentner, A. R. 2005, ApJ 634,
344

\bibitem[Gualandris \& Portegies Zwart(2007)]{t15_gual07} 
Gualandris, A., \& Portegies Zwart, S.\ 2007, \mnras, 376, L29 

\bibitem[Gvaramadze et al.(2007)]{t15_gvar07} Gvaramadze, V.~V., 
Gualandris, A., \& Portegies Zwart, S.\ 2007, arXiv:0712.4230 




\bibitem[Heber et al.(2000)]{t15_heber00}
   Heber,~U., Reid,~I.~N., Werner,~K. 
   2000, \aap, 363, 198 


\bibitem[Hills(1988)]{t15_hills88} Hills, J. G. 1988, Nature 331, 687

\bibitem[Hirsch et al.(2005)]{t15_hirsch05} Hirsch, H. A., Heber, U., O'Toole, 
S. J., \& Bresolin, F. 2005, A\&A 444, L61



\bibitem[O'Toole \& Heber(2006)]{t15_otoole06} O'Toole, S.~J., \& 
Heber, U.\ 2006, \aap, 452, 579 



\bibitem[Schaller et al.(1992)]{t15_schaller92} Schaller, G., 
Schaerer, D., Meynet, G., \& Maeder, A.\ 1992, \aaps, 96, 269 

\bibitem[Str\"{o}er et al.(2007)]{t15_stroer07}
Str\"{o}er, A., Heber, U., Lisker, T., et al.\ 2007, A\&A, 462, 269

\bibitem[Yu \& Tremaine(2003)]{t15_yu03} Yu, Q., \& Tremaine, S. 2003, ApJ 599, 112

\end{thebibliography}
\end{document}